\documentclass[aps,prb,reprint,amssymb,superscriptaddress]{revtex4-1}

\usepackage{amsmath,amssymb,color}
\usepackage{bm}
\usepackage{hyperref}
\usepackage{graphicx}

\begin{document}

\title{Ellipticity and Dissipation Effects in Magnon Spin Valves}

\author{Jiansen Zheng}
\affiliation{Institute for Theoretical Physics and Center for Extreme Matter and Emergent Phenomena,
Utrecht University, Leuvenlaan 4, 3584 CE Utrecht, The Netherlands}

\author{Andreas R\"{u}ckriegel}
\affiliation{Institute for Theoretical Physics and Center for Extreme Matter and Emergent Phenomena,
Utrecht University, Leuvenlaan 4, 3584 CE Utrecht, The Netherlands}

\author{Scott A. Bender}
\affiliation{Institute for Theoretical Physics and Center for Extreme Matter and Emergent Phenomena,
Utrecht University, Leuvenlaan 4, 3584 CE Utrecht, The Netherlands}

\author{Rembert A. Duine}
\affiliation{Institute for Theoretical Physics and Center for Extreme Matter and Emergent Phenomena,
Utrecht University, Leuvenlaan 4, 3584 CE Utrecht, The Netherlands}
\affiliation{Center for Quantum Spintronics, Department of Physics, Norwegian University of Science and Technology, 
NO-7491 Trondheim, Norway}
\affiliation{Department of Applied Physics, Eindhoven University of Technology,
P.O. Box 513, 5600 MB Eindhoven, The Netherlands}

\date{March 2, 2020}

\begin{abstract}
We consider alignment-dependent spin and heat transport across a magnon spin valve in the tunneling regime, i.e., a junction consisting of two weakly coupled ferromagnetic insulators. We determine the difference in spin and heat conductance between the parallel and antiparallel configuration of the magnetization direction. The dependence of these conductances on both the Gilbert damping and ellipticity is studied. We find that both magnon ellipticity and dissipation open channels for magnons to tunnel through in the antiparallel configuration. Our results highlight an important difference between electronic and magnon spin transport in spin-valve structures and may be important for the development of devices based on magnetic insulators. 
\end{abstract}

\maketitle

\section{Introduction}

Spintronics based on spin-polarized charge currents has led to a boost in information storage technology with the discovery of giant magnetoresistance (GMR) in antiferromagnetically coupled $\rm Fe/Cr$ superlattices \cite{baibich1988giant,dieny1994giant}. GMR arises from the spin-dependent transmission of the conduction electron. 
Magnons, the quanta of collective excitations in magnetically ordered systems, 
can carry spin current  in magnetic insulators in the absence of any charge current, e.g., in the spin Seebeck effect \cite{uchida2010observation}, where the magnons are driven by a thermal bias, or in nonlocal setups in which the magnons are biased electrically using the spin Hall effect in adjacent normal metals \cite{cornelissen2015long}. Magnon spin transport is promising, for example, to improve the power efficiency of logic devices \cite{chumak2015magnon,klingler2015spin,chumak2014magnon} 
and for neuromorhpic computing \cite{torrejon2017neuromorphic,bracher2018analog}.

To find analogies of GMR in magnon spin transport, spin-valve structures that encompass magnetic insulators 
have recently been studied experimentally and theoretically \cite{cramer2018magnon,hamalainen2018programmable,guo2018magnon}. 
Wu {\it et al.} have observed that the spin Seebeck effect of a heterostructure made of two ferromagnetic insulators, namely yttrium iron garnet (YIG),  separated by  a nonmagnetic heavy metal layer, depends on the relative orientation of the  magnetizations of the two magnetic insulators \cite{wu2018magnon}. The difference in spin Seebeck signal between parallel and antiparallel configurations is observed to decrease significantly as the temperature is lowered. In a recent theoretical  work, a Green's function formalism for magnon tunneling driven by a temperature bias across a ferromagnetic junction has been developed and applied to compute the diode properties of the tunneling magnon current \cite{ren2013theory}. A key aspect of this study is the inclusion of magnon-magnon interactions that are exploited for the rectification and negative differential spin Seebeck effects.
Furthermore, a tunable spin Seebeck diode based on a magnetic junction structure in which the tunneling spin current can be turned on and off by controlling the magnetization orientation has also been theoretically proposed \cite{diasuke2018tunable}. 

\begin{figure}[!htb]
\includegraphics[width=0.35\textwidth]{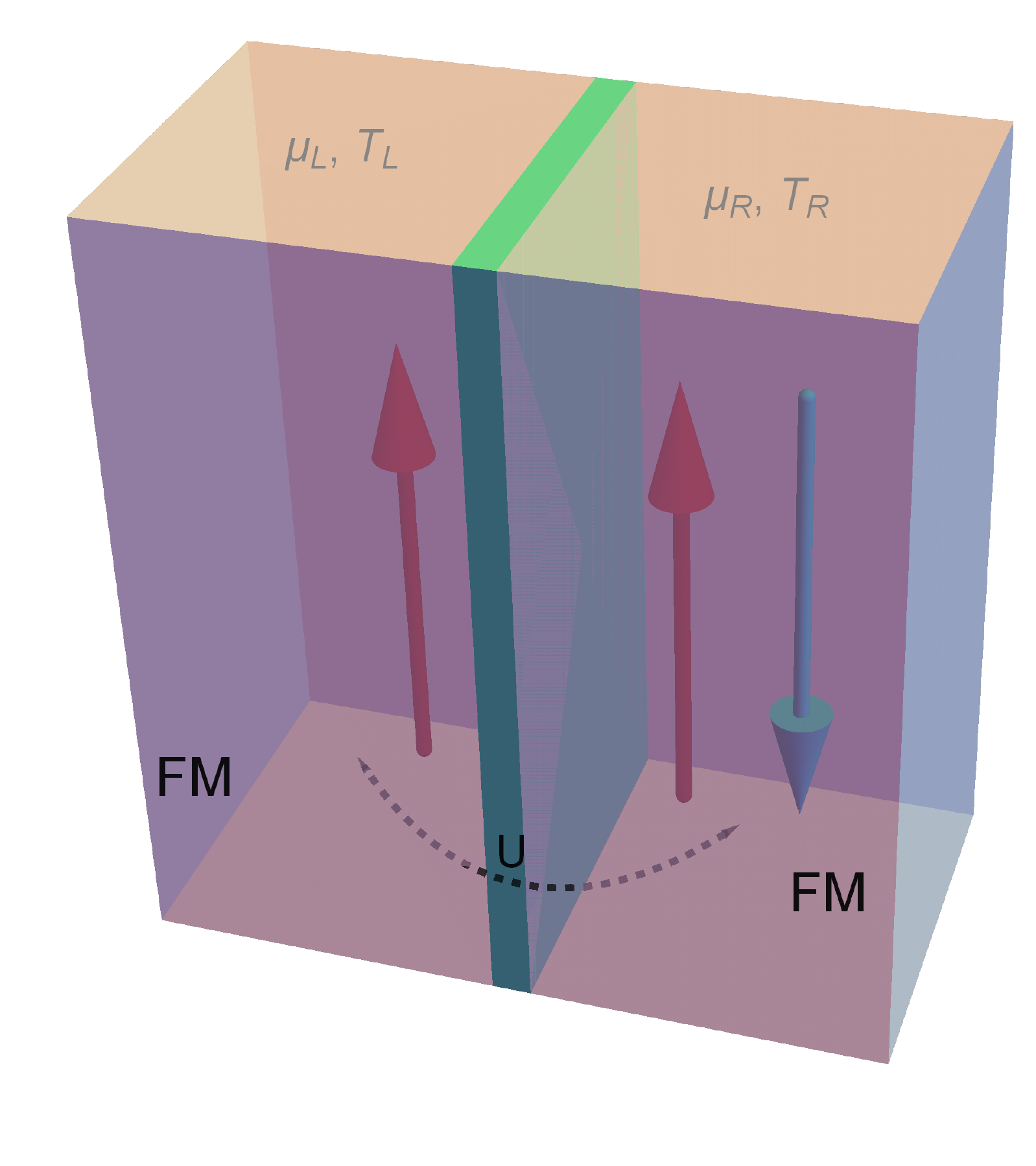}
\label{fig:mtj_setup}
\caption{Illustration of a magnetic junction consisting of two ferromagnetic insulators, which interact with each other with weak exchange coupling $U$. The left (right) insulator has temperature $T_L (T_R)$ and spin accumulation $\mu_L (\mu_R)$. The layer in green is a nonmagnetic insulator and is thin enough for the magnons to tunnel across it. The left insulator is the fixed layer, with the red arrow denoting an up spin, while the right insulator is a free layer in which the magnetization can be tuned from a parallel configuration (denoted with the red arrow) to an anti-parallel configuration (denoted with the blue arrow).
}
\end{figure}

In this work, we study the alignment dependence of magnon heat and spin transport across a heterostructure consisting of two ferromagnetic insulators that are weakly exchange coupled, e.g., by a nonmagnetic spacer layer that mediates exchange interactions. The setup we consider is illustrated in Fig.~\ref{fig:mtj_setup}. 
The ferromagnetic insulators act as reservoirs for magnons. The magnons can be coherently driven by ferromagnetic resonance, or incoherently generated with an electrical or thermal bias using adjacent normal metals \cite{cornelissen2016temperature}. 
We focus on the effect of the ellipticity of the magnetization precession, which is usually caused by anisotropies, 
and also on the effects of dissipation that we parametrize with a Gilbert damping constant. 
The latter is a phenomenological parameter that characterizes the decay of magnons. 
The ellipticity of precession has been shown to strongly affect the parametric excitation of magnons in ferromagnetic resonance experiments \cite{langner2017damping}, 
and plays a role in Rayleigh-Jeans condensation of pumped magnons in thin-film ferromagnets \cite{ruckriegel2015rayleigh}. Moreover, at the quantum-mechanical level, the ellipticity leads to squeezed ground states and, in the case of antiferromagnets, entanglement between different sublattice magnons \cite{kamra2019fock,Zou2019entanglement}. 
Meanwhile, a low Gilbert damping has also been demonstrated to enable long-distance spin transport in magnetic insulators such as yttrium iron garnet \cite{cornelissen2015long}. 
Neither the influence of damping nor ellipticity has, however, been considered for magnon tunneling in the insulating spin valve structure considered here.

In our setup, magnons tunnel between the ferromagnets due to the weak exchange coupling and carry heat and spin currents in response to applied temperature or magnon chemical potential differences. For circularly polarized magnons, conservation of spin forbids magnon tunneling in the antiparallel configuration. We find that anisotropies and dissipation, both of which break spin conservation, lead to tunneling currents even in the antiparallel configuration. The difference between circular and elliptical case has no straightforward analog in electronic spin valves, as in these latter systems the spin of the electrons at the Fermi level is usually approximately conserved, and tunneling current is determined by their spin-dependent density of states. Our result therefore shows that the difference between magnon currents in the parallel and antiparallel configuration is governed by different physics than for metallic structures, which may be useful in designing devices that exploit the tunability of the tunneling current.

The remainder of the paper is organized as follows: 
In Sec.~\ref{sec:Hamiltonian},
we introduce the model that we use for our setup. 
The magnon tunneling currents in the presence of both precession ellipticity and Gilbert damping 
for both parallel and antiparallel configurations
are calculated in Sec.~\ref{sec:currents}.
In Sec.~\ref{sec:results}, we numerically calculate the tunneling conductances
and discuss their dependencies on the ellipticity and dissipation.
Section \ref{sec:conclusions} summarizes our main findings and conclusions.
Lastly, the Appendix outlines the derivation of the tunneling currents using rate equations.

\section{Model Hamiltonian}

\label{sec:Hamiltonian}

%
%
%
%

\subsection{Lead Hamiltonians}

\label{sec:H_X}

The magnetization dynamics in the bulk of each insulating lead is modeled by the Hamiltonian
\begin{align}
{\cal H}_X =
& - \frac{1}{2} \sum_{ij} J_{X,ij} \bm{S}_{X,i} \cdot \bm{S}_{X,j} 
- \hbar \gamma_X \mu_0 H_X \sum_i S_{X,i}^z
\nonumber\\
&- \frac{1}{2} \sum_i \left[ K_{X,x} \left( S_{X,i}^x \right)^2 + K_{X,y} \left( S_{X,i}^y \right)^2 \right] , 
\label{eq:H_X}
\end{align}
where $X=L/R$ denotes the left/right lead and $i,j$ label the lattice sites. Here,
$J_{X,ij}$ are nearest-neighbor exchange interactions with strength $J_X > 0$,
whereas $K_{X,x}$ and $K_{X,y}$ are anisotropy constants.
Lastly, $\mu_0 H_X$ are the magnetic fields in the bulk of the leads with gyromagnetic ratio $\gamma_X$.

\begin{figure}
\includegraphics[width=0.7\columnwidth]{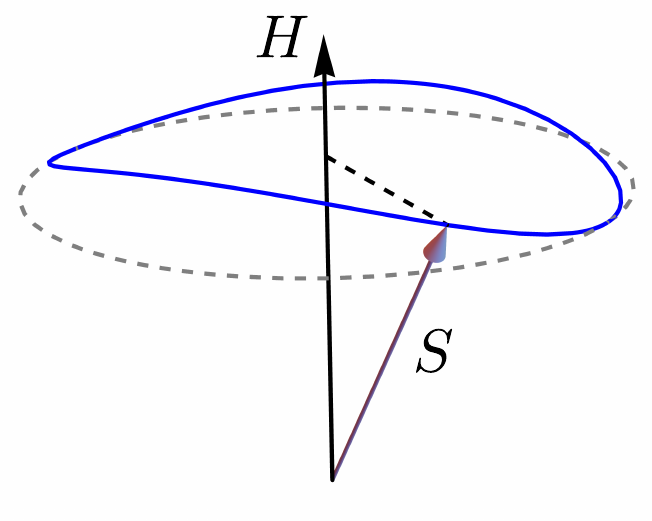}
\caption{ \label{fig:ellipse}
Semiclassical depiction of a spin $\bm{S}$ that precesses elliptically around a magnetic field $\bm{H}$.
The solid blue line is the elliptical precession,
whereas the dashed gray line corresponds to a circular precession.
Because the length of the spin is conserved, its projection onto the direction of the magnetic field is not constant
during the elliptical precession. 
Therefore, this spin projection is no longer a good quantum number for elliptical magnons.
}
\end{figure}

The spin operators $\bm{S}_{X,i}$ are bosonized via a Holstein-Primakoff transformation \cite{holstein1940field}.
For $H_X>0$ we assume that the magnetic order parameter points in $z$ direction, 
so that 
\begin{subequations} \label{eq:HP_up}
\begin{align}
S_{X,i}^+ =& S_{X,i}^x + i S_{X,i}^y = \sqrt{2S_X} \left[ b_{X,i} + {\cal O}(S_X^{-1}) \right], \\
S_{X,i}^z =& S_X - b_{X,i}^\dagger b_{X,i} ,
\end{align}
\end{subequations}
where $S_X$ is the spin quantum number of the magnetic moments in lead $X$,
and $b_{X,i}$ and $b_{X,i}^\dagger$ are the magnon annihilation and creation operators
that satisfy the bosonic commutation relations $ [b_{X,i},b_{X',j}^\dagger] = \delta_{X,X'} \delta_{i,j} $.
Conversely, for $H_X<0$
we assume that the magnetic order parameter points in $-z$ direction
and apply the following Holstein-Primakoff transformation:
\begin{subequations} \label{eq:HP_down}
\begin{align}
S_{X,i}^+ =& S_{X,i}^x + i S_{X,i}^y = \sqrt{2S_X} \left[ b_{X,i}^\dagger + {\cal O}(S_X^{-1}) \right], \\
S_{X,i}^z =& - S_X + b_{X,i}^\dagger b_{X,i} .
\end{align}
\end{subequations}
In the bulk of each lead,
we may expand the magnon operators in a Fourier series as
\begin{equation} \label{eq:Fourier}
b_{X,i} = \frac{1}{\sqrt{N_X}} \sum_{\bm{k}} e^{i\bm{k}\cdot\bm{R}_i} b_{X,\bm{k}} ,
\end{equation}
where $N_X$ denotes the number of magnetic moments in the lead $X$.
Then the spin Hamiltonian (\ref{eq:H_X}) becomes  
\begin{equation} \label{eq:H_X_boson}
{\cal H}_X = \sum_{\bm{k}} \left[
A_{X,\bm{k}} b_{X,\bm{k}}^\dagger b_{X,\bm{k}} 
+ \frac{B_X}{2} \left( b_{X,\bm{k}}^\dagger b_{X,-\bm{k}}^\dagger + \textrm{h.c.} \right)
\right] ,
\end{equation}
where we dropped a constant contribution to the ground state energy as well as 
${\cal O}(S_X^{-1/2})$ corrections containing higher powers of the magnon operators.
The coefficients of the Hamiltonian (\ref{eq:H_X_boson}) are given by
\begin{subequations}
\begin{align}
A_{X,\bm{k}} =& \hbar \gamma_X \mu_0 |H_X| + J_X S_X a_X^2 \bm{k}^2 - \frac{S_X}{2} \left( K_{X,x} + K_{X,y} \right) , 
\label{eq:A_longWavelength}
\\
B_X =& - \frac{S_X}{2} \left( K_{X,x} - K_{X,y} \right) ,
\end{align}
\end{subequations}
regardless of whether we assume $H_X>0$ or $H_X<0$ 
and employ the respective Holstein-Primakoff transformation (\ref{eq:HP_up}) or (\ref{eq:HP_down}).
In writing down Eq.~(\ref{eq:A_longWavelength}),
we furthermore assumed that only long-wavelength magnons with $a_X|\bm{k}|\ll 1$ are relevant,
where $a_X$ is the lattice constant of lead $X$.
The quadratic magnon Hamiltonian (\ref{eq:H_X_boson}) is diagonalized
via a Bogoliubov transformation:
\begin{equation} \label{eq:Bogoliubov}
\left( \begin{matrix}
b_{X,\bm{k}} \\ b_{X,-\bm{k}}^\dagger 
\end{matrix} \right) =
\left( \begin{matrix}
u_{X,\bm{k}} & -v_{X,\bm{k}} \\ -v_{X,\bm{k}} & u_{X,\bm{k}}
\end{matrix} \right)
\left( \begin{matrix}
\beta_{X,\bm{k}} \\ \beta_{X,-\bm{k}}^\dagger 
\end{matrix} \right) ,
\end{equation}
where
\begin{subequations} \label{eq:u_v}
\begin{align}
u_{X,\bm{k}} =& \sqrt{ \frac{ A_{X,\bm{k}} + E_{X,\bm{k}} }{2 E_{X,\bm{k}}} } , \\
v_{X,\bm{k}} =& \frac{B_X}{|B_X|} \sqrt{ \frac{ A_{X,\bm{k}} - E_{X,\bm{k}} }{2 E_{X,\bm{k}}} } ,
\end{align}
\end{subequations}
where
\begin{equation} \label{eq:dispersion}
E_{X,\bm{k}} = \sqrt{ \left( A_{X,\bm{k}} + B_X \right) \left( A_{X,\bm{k}} - B_X \right) } .
\end{equation}
The operators $\beta_{X,\bm{k}}$ and $\beta_{X,\bm{k}}^\dagger$ create and destroy
Bogoliubov quasiparticles and obey the bosonic commutation relations 
$ [ \beta_{X,\bm{k}} , \beta_{X',\bm{k}'}^\dagger ] = \delta_{X,X'} \delta_{\bm{k},\bm{k}'}  $.
Semiclassically, a Bogoliubov quasiparticle created by $\beta_{X,\bm{k}}^\dagger$ corresponds to an
elliptical spin wave;
in contrast, a magnon created by $b_{X,\bm{k}}^\dagger$ corresponds to a circular spin wave.
The Bogoliubov quasiparticles are often also referred to as magnons,
or as elliptical or squeezed magnons \cite{kamra2019fock,Zou2019entanglement}.
For an elliptical spin wave, the $z$ component of the spin is not conserved and hence not a good quantum number,
unlike for a circular spin wave.
This is illustrated semiclassically in Fig.~\ref{fig:ellipse}.
With the Bogoliubov transformation (\ref{eq:Bogoliubov}), 
the magnon Hamiltonian (\ref{eq:H_X_boson}) becomes
\begin{equation} \label{eq:H_X_Bogoliubov}
{\cal H}_X = \sum_{\bm{k}} \left[ E_{X,\bm{k}} \beta_{X,\bm{k}}^\dagger \beta_{X,\bm{k}}
+ \frac{1}{2} \left( E_{X,\bm{k}} - A_{X,\bm{k}} \right) \right] .
\end{equation}
Note that this Hamiltonian is only valid as long as the quasiparticle dispersion (\ref{eq:dispersion}) is real,
i.e., for
$ \hbar \gamma_X \mu_0 |H_X| > S K_{X,x} , S K_{X,y} $.
If this is not satisfied,
our original assumption that the magnetic order points in $\pm z$ direction is not correct
and we have to expand around a different ground state.
However, for the remainder of this work
we will assume that $ \hbar \gamma_X \mu_0 |H_X| > S K_{X,x} , S K_{X,y} $,
so that the quasiparticle Hamiltonian (\ref{eq:H_X_Bogoliubov}) is stable.

\subsection{Tunneling Hamiltonian}

The tunneling between the leads is facilitated by a lead-lead exchange interaction of the form
\begin{equation} \label{eq:H_T}
{\cal H}_T = - \sum_{ij} U_{ij} \bm{S}_{L,i} \cdot \bm{S}_{R,j} ,
\end{equation}
where the exchange coupling $U_{ij}$ is assumed to be small compared to the bulk energy scales and 
only finite for $i,j$ close to the interface. 
The microscopic origin of such an interaction could either be direct exchange mediated by the conduction electrons in the normal metal 
or an indirect superexchange interaction via the ions in the non-magnetic spacer layer.

\subsubsection{Parallel Configuration}

\label{sec:H_T_parallel}

In the parallel configuration,
we take the form (\ref{eq:HP_up}) for both leads.
Then the tunneling Hamiltonian (\ref{eq:H_T}) becomes
\begin{equation}
\label{eq:circmagnonspar}
{\cal H}_T^P = - \sqrt{S_L S_R} \sum_{ij} U_{ij} \left(
b_{L,i}^\dagger b_{R,j} + b_{R,j}^\dagger b_{L,i} \right) ,
\end{equation}
where we dropped constants and higher order magnon corrections as before,
as well as an on-site energy shift for the magnons in each lead.
This is justified by our assumption that the lead-lead exchange is small
compared to the bulk energy scales.
After applying both Fourier and Bogoliubov transformations,
Eqs.~(\ref{eq:Fourier}) and (\ref{eq:Bogoliubov}) respectively,
we find
\begin{align}
{\cal H}_T^P =
& 
- \sqrt{ \frac{S_L S_R}{N_L N_R} } \sum_{\bm{k}\bm{k}'} 
\nonumber\\
&
%
%
%
%
\times
\biggl(
V_{\bm{k},\bm{k}'}^{(n)} \beta_{L,\bm{k}}^\dagger \beta_{R,\bm{k}'} -
V_{\bm{k},\bm{k}'}^{(a)} \beta_{L,\bm{k}}^\dagger \beta_{R,-\bm{k}'}^\dagger
+ \textrm{h.c.}
\biggr) ,
\end{align}
where 
\begin{subequations}
\begin{align}
V_{\bm{k},\bm{k}'}^{(n)} =
&
U_{\bm{k},-\bm{k}'} \left( u_{L,\bm{k}} u_{R,\bm{k}'} + v_{L,\bm{k}} v_{R,\bm{k}'} \right), 
\\
V_{\bm{k},\bm{k}'}^{(a)} =
&
U_{\bm{k},-\bm{k}'} \left( u_{L,\bm{k}} v_{R,\bm{k}'} + v_{L,\bm{k}} u_{R,\bm{k}'} \right)
\label{eq:Va}
\end{align}
\end{subequations}
are the normal ($n$) and anomalous ($a$) tunneling amplitudes,
with the Fourier transform of the lead-lead exchange coupling 
$ U_{\bm{k},\bm{k}'} = \left( U_{-\bm{k},-\bm{k}'} \right)^*
= \sum_{i\in L} \sum_{j\in R} e^{-i\bm{k}\cdot\bm{R}_i -i\bm{k}'\cdot\bm{R}_j} U_{ij} $.
Note that the anomalous coupling (\ref{eq:Va}) is only finite when the magnon ellipticity is finite,
leading to qualitatively new physics in this case.

\subsubsection{Antiparallel Configuration}

In the antiparallel configuration,
we take the Holstein-Primakoff transformations (\ref{eq:HP_up}) for the left
and (\ref{eq:HP_down}) for the right lead,
yielding
\begin{equation}
\label{eq:circmagnonsap}
{\cal H}_T^{AP} = - \sqrt{S_L S_R} \sum_{ij} U_{ij} \left(
b_{L,i}^\dagger b_{R,j}^\dagger + b_{L,i} b_{R,j} \right) ,
\end{equation}
within the same approximations as for the parallel configuration considered in the preceding Sec.~\ref{sec:H_T_parallel}.
As before, we apply the Fourier and Bogoliubov transformations
given, respectively, in Eqs.~(\ref{eq:Fourier}) and (\ref{eq:Bogoliubov}) 
to obtain
\begin{align}
{\cal H}_T^{AP} =
& 
- \sqrt{ \frac{S_L S_R}{N_L N_R} } \sum_{\bm{k}\bm{k}'} 
\nonumber\\
&
%
%
%
%
\times
\biggl(
V_{\bm{k},\bm{k}'}^{(n)} \beta_{L,\bm{k}}^\dagger \beta_{R,-\bm{k}'}^\dagger -
V_{\bm{k},\bm{k}'}^{(a)} \beta_{L,\bm{k}}^\dagger \beta_{R,\bm{k}'}
+ \textrm{h.c.}
\biggr)
\end{align}
Note that from the comparison of the magnon tunneling Hamiltonian between the parallel and antiparallel case, Eq.~(\ref{eq:circmagnonspar}) and Eq.~(\ref{eq:circmagnonsap}), respectively, it is clear that these two situations differ qualitatively. In the parallel case, one deals with a tunneling Hamiltonian that is also encountered in the electron transport, whereas in the antiparallel case, the tunneling corresponds to creation or destruction of a pair of circular magnons.
We also stress that the magnon ellipticity, being related to the breaking of magnon number, i.e., spin conservation, has no analog in electronic systems,
where the electron number is always conserved.
Therefore, the effects discussed here have no direct analog in electronic valves.
From Eq.~(\ref{eq:circmagnonsap}) it is furthermore obvious that there is no spin transport in the antiparallel configuration 
without breaking of the total spin conservation, 
either by anisotropies or damping.
In this respect, the undamped, circular magnon spin valve resembles a half-metallic system 
that only transmits spin when the magnetizations of the two magnets are aligned parallel.

\section{Tunneling Currents}

\label{sec:currents}

The tunneling currents can be obtained from the 
rate equations for the distribution function of the 
Bogoliubov quasiparticles in each lead,
\begin{equation} \label{eq:nk}
n_{X,\bm{k}} = \left\langle \beta_{X,\bm{k}}^\dagger \beta_{X,\bm{k}} \right\rangle
= f_B\left( \frac{ E_{X,\bm{k}} - \mu_X }{ k_B T_X } \right)
\end{equation}
where $f_B(x) = 1/ (e^x-1)$ is the Bose function, 
and the second equality holds in a steady state 
in which lead $X$ is kept at temperature $T_X$ and chemical potential $\mu_X$.
To allow for finite damping in each lead, 
we recast the steady state distribution function (\ref{eq:nk}) as 
\begin{equation} \label{eq:nk_int}
n_{X,\bm{k}} 
= \int_{-\infty}^\infty d\epsilon\, \delta\left( \epsilon - E_{X,\bm{k}} \right) 
f_B\left( \frac{ \epsilon - \mu_X }{ k_B T_X } \right) .
\end{equation}
Within the Gilbert damping phenomenology,
we may then add dissipation by broadening the Dirac distributions according to \cite{ren2013theory,zheng2017green}
\begin{equation} \label{eq:dissipation}
\delta\left( \epsilon - E_{X,\bm{k}} \right) \to A\left( \epsilon - E_{X,\bm{k}} \right) 
\equiv \frac{1}{\pi} \frac{ \alpha \epsilon }{ \left( \epsilon - E_{X,\bm{k}} \right)^2 + \left( \alpha \epsilon \right)^2 } ,
\end{equation}
where $\alpha$ is the bulk Gilbert damping parameter. 

Details of the derivation of the tunneling currents
from kinetic equations for the quasiparticle distribution functions
can be found in the Appendix; 
here we only state the results.
Labeling the parallel/antiparallel configurations with 
$ Y = P / AP $,
we find the following expressions for the energy current 
\begin{align}
I_E^Y =
&
\frac{2\pi}{\hbar} \int_{-\infty}^\infty d\epsilon\, \epsilon
\nonumber\\
&
\times \Biggl\{
D_E^Y(\epsilon) 
\left[ f_B\left( \frac{ \epsilon - \mu_L }{ k_B T_L } \right) - f_B\left( \frac{ \epsilon - \mu_R }{ k_B T_R } \right) \right]
\nonumber\\
&
\phantom{ \Biggl\{ }
+ \tilde{D}_E^Y(\epsilon) 
\left[ f_B\left( \frac{ \epsilon - \mu_L }{ k_B T_L } \right) - f_B\left( \frac{ \epsilon + \mu_R }{ k_B T_R } \right) \right]
\Biggr\} ,
\label{eq:I_E}
\end{align}
and the spin current
\begin{align} 
I_S^Y =
&
2\pi \int_{-\infty}^\infty d\epsilon\,
\nonumber\\
&
\times \Biggl\{
D_S^Y(\epsilon) 
\left[ f_B\left( \frac{ \epsilon - \mu_L }{ k_B T_L } \right) - f_B\left( \frac{ \epsilon - \mu_R }{ k_B T_R } \right) \right]
\nonumber\\
&
\phantom{ \Biggl\{ }
+ \tilde{D}_S^Y(\epsilon) 
\left[ f_B\left( \frac{ \epsilon - \mu_L }{ k_B T_L } \right) - f_B\left( \frac{ \epsilon + \mu_R }{ k_B T_R } \right) \right]
\Biggr\} ,
\label{eq:I_S}
\end{align}
flowing from the left to the right lead.
Here, $D_{E/S}^{P/AP}(\epsilon)$ are the normal tunneling densities of state,
explicitly given by
\begin{align} \label{eq:D_P_AP}
\left\{ \begin{matrix}
D_E^{P/AP}(\epsilon) \\
D_S^{P/AP}(\epsilon)
\end{matrix} \right\}
=
&
\frac{S_L S_R}{N_R N_L} \sum_{\bm{k}\bm{k}'} \left| V_{\bm{k},\bm{k}'}^{(n/a)} \right|^2
\left\{ \begin{matrix}
1 \\
\left( u_{R,\bm{k}'}^2 + v_{R,\bm{k}'}^2 \right)
\end{matrix} \right\}
\nonumber\\
&
\times
A\left( \epsilon - E_{L,\bm{k}} \right) A\left( \epsilon - E_{R,\bm{k}'} \right) .
\end{align}
Note that their contributions to currents (\ref{eq:I_E}) and (\ref{eq:I_S})
vanish if both leads are at the same temperature and chemical potential.
On the other hand,
$\tilde{D}_{E/S}^{P/AP}(\epsilon)$ are anomalous tunneling densities of state
that arise because the Gilbert damping breaks the number conservation
of the Bogoliubov quasiparticles.
Hence it gives rise to a spin current even when both leads are at the same temperature and chemical potential;
it vanishes only if both leads are in true thermal equilibrium at the same temperature and vanishing chemical potential.
These anomalous tunneling densities of state are
\begin{align} \label{eq:tildeD_P_AP}
\left\{ \begin{matrix}
\tilde{D}_E^{P/AP}(\epsilon) \\
\tilde{D}_S^{P/AP}(\epsilon)
\end{matrix} \right\}
=
&
\frac{S_L S_R}{N_R N_L} \sum_{\bm{k}\bm{k}'} \left| V_{\bm{k},\bm{k}'}^{(a/n)} \right|^2
\left\{ \begin{matrix}
1 \\
\left( u_{R,\bm{k}'}^2 + v_{R,\bm{k}'}^2 \right)
\end{matrix} \right\}
\nonumber\\
&
\times
A\left( \epsilon - E_{L,\bm{k}} \right) A\left( \epsilon + E_{R,\bm{k}'} \right) .
\end{align}

\begin{figure}
\includegraphics[width=0.9\columnwidth]{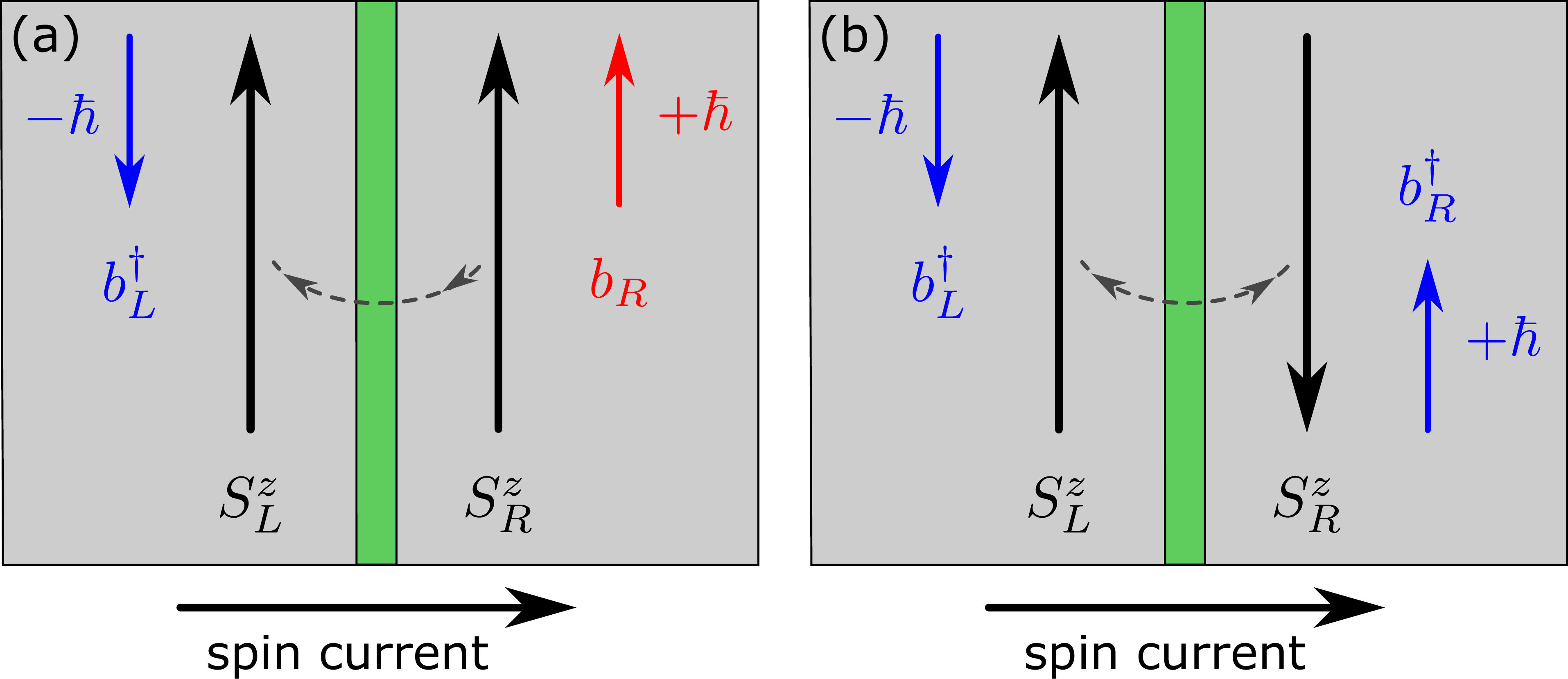}
\caption{ \label{fig:TunnelingCircular}
Tunneling processes allowed by spin conservation for circular magnons.
(a) Hopping of a magnon carrying spin $-\hbar$ from the right to the left lead in the parallel configuration. 
The inverse process of hopping from left to right lead is also possible.
(b) Pair creation of a magnon carrying spin $-\hbar$ in the left lead 
and a magnon carrying spin $+\hbar$ in the right lead in the antiparallel configuration.
The inverse process of pair annihilation is also allowed.
However, while allowed by spin conservation,
both pair creation and annihilation processes are forbidden by energy conservation
if there is no dissipation. 
The spin currents in (a) and (b) are polarized in the $z$ direction.
For the inverse processes, the spin currents flow in the opposite direction.
The blue (red) arrows indicate the spin change associated with the creation (annihilation) of a circular magnon.
}
\end{figure}

It is instructive to consider the limit of conserved quasiparticles ($\alpha=0^+$)
as well as the limit of circular magnons in more detail.
If there is no dissipation,
the anomalous contributions to the currents vanish 
because energy conservation strictly demands $E_{L,\bm{k}} + E_{R,\bm{k}'} = 0$,
which can never be satisfied since both of these energies are positive.
Finite damping softens this restriction by allowing energy (and spin) transfer to a thermal bath,
thereby opening up another channel for energy and spin transfer between the leads.
In the limit of circular magnons, 
i.e., when there are no anisotropies that break rotation symmetry around the $z$ direction,
we may set $u_{X,\bm{k}} = 1$ and $v_{X,\bm{k}} = 0$;
see Eqs.~(\ref{eq:u_v}). 
Then $\tilde{D}_{E/S}^P(\epsilon)=0=D_{E/S}^{AP}(\epsilon)$.
This reflects the conservation of the total spin $S_L^z + S_R^z$ in the absence of anisotropies.
In the parallel configuration, tunneling is in this case only allowed
for the process in which a magnon carrying spin $-\hbar$ is destroyed in one lead 
and another magnon carrying spin $-\hbar$ is created in the other lead.
Conversely, in the antiparallel configuration 
magnons carry spin $-\hbar$ in the left lead and $+\hbar$ in the right lead.
Thus spin conservation only allows anomalous processes in which
magnon pairs in the left and right lead are simultaneously destroyed or created.
As this process violates energy conservation,
it is only possible in the presence of dissipation.
Therefore, there are no energy and spin currents in the antiparallel configuration without 
either damping (enabling pair creation/annihilation processes)
or breaking of spin conservation (enabling normal hopping).
This is further illustrated in Fig.~\ref{fig:TunnelingCircular}.

\subsection{Tunneling Conductances}

If the biasing is sufficiently small,
i.e.,
$ \Delta T = T_L - T_R \ll T $,
where
$ T = \frac{1}{2} \left( T_L + T_R \right) $
is the average temperature,
and 
$ \mu_{L/R}  \ll E_{L/R,\bm{k}=0} $,
we may linearize the Bose functions appearing in the currents (\ref{eq:I_E}) and (\ref{eq:I_S}),
yielding
\begin{align}
I_E^Y =& \kappa^Y \Delta T + \Pi^Y \left( \mu_L - \mu_R \right) + \gamma_E^Y \left( \mu_L + \mu_R \right), \\
I_S^Y =& L^Y \Delta T + \sigma^Y \left( \mu_L - \mu_R \right) + \gamma_S^Y \left( \mu_L + \mu_R \right).
\end{align}
Here,
\begin{subequations} \label{eq:conductances}
\begin{equation}
\kappa^Y =
\frac{\pi}{2 \hbar k_B T^2} \int_{-\infty}^\infty d\epsilon\,
\frac{ \epsilon^2 }{ \sinh^2\left( \frac{ \epsilon }{ 2 k_B T } \right) }
\left[ D_E^Y(\epsilon) + \tilde{D}_E^Y(\epsilon) \right] ,
\end{equation}
is the thermal conductance,
\begin{equation} \label{eq:Seebeck}
L^Y = 
\frac{\pi}{2 k_B T^2} \int_{-\infty}^\infty d\epsilon\,
\frac{ \epsilon }{ \sinh^2\left( \frac{ \epsilon }{ 2 k_B T } \right) }
\left[ D_S^Y(\epsilon) + \tilde{D}_S^Y(\epsilon) \right] ,
\end{equation}
is the spin Seebeck conductance,
\begin{equation} \label{eq:Peltier}
\Pi^Y =
\frac{\pi}{2 \hbar k_B T} \int_{-\infty}^\infty d\epsilon\,
\frac{ \epsilon  }{ \sinh^2\left( \frac{ \epsilon }{ 2 k_B T } \right) }
D_E^Y(\epsilon) ,
\end{equation}
is the spin Peltier conductance,
and
\begin{equation}
\sigma^Y =
\frac{\pi}{2 k_B T} \int_{-\infty}^\infty d\epsilon\,
\frac{ 1 }{ \sinh^2\left( \frac{ \epsilon }{ 2 k_B T } \right) }
D_S^Y(\epsilon) ,
\end{equation}
is the spin conductance.
%
%
%
%
%
%
%
%
%
%
Lastly,
\begin{align}
\gamma_E^Y =
& 
\frac{\pi}{2 \hbar k_B T} \int_{-\infty}^\infty d\epsilon\,
\frac{ \epsilon  }{ \sinh^2\left( \frac{ \epsilon }{ 2 k_B T } \right) }
\tilde{D}_E^Y(\epsilon) ,
\label{eq:gamma_E}
\\
\gamma_S^Y =
& 
\frac{\pi}{2 k_B T} \int_{-\infty}^\infty d\epsilon\,
\frac{ 1 }{ \sinh^2\left( \frac{ \epsilon }{ 2 k_B T } \right) }
\tilde{D}_S^Y(\epsilon) 
\end{align}
\end{subequations}
are the additional energy and spin loss or gain terms arising because the finite damping breaks
the number conservation of Bogoliubov quasiparticles.
Since they do not vanish when both leads are mutually equilibrated, 
these terms are not part of the transport current 
and should rather be identified
with the spin and energy lost to or gained from the thermal bath that provides the dissipation,
which is ultimately the crystal lattice.  
Microscopically, they correspond to the simultaneous creation or annihilation of a magnon in the left and a magnon in the right lead;
the required energy and angular momentum is provided by the lattice. 
Therefore, these terms describe energy and spin currents flowing from the lattice to both leads, 
instead of currents flowing from one lead to the other.
Because the total angular momentum of spins and lattice is conserved,
this additional spin transfer should be experimentally detectable
as torques on the whole sample.

Note also that the spin Seebeck and Peltier conductances, 
Eqs.~(\ref{eq:Seebeck}) and (\ref{eq:Peltier}), respectively,
are not Onsager reciprocals of each other,
$\hbar \Pi^Y \neq T L^Y$. 
There are two independent reasons for this: 
the breaking of time-reversal symmetry by the dissipation
and the breaking of spin conservation by the anisotropies.
While the former opens up an new channel for bath-assisted energy transfer,
namely the pair creation/annihilation processes contained in $\tilde{D}_{E/S}^Y(\epsilon)$,
the latter allows for changes in spin without accompanying changes in energy,
resulting in $D_{E}^Y(\epsilon) \neq D_{S}^Y(\epsilon)$ and
$\tilde{D}_{E}^Y(\epsilon) \neq \tilde{D}_{S}^Y(\epsilon)$.

\section{Numerical Results and Discussion}

\label{sec:results}

In realistic systems, the
interfaces between layers of different materials are usually rough.
Such rough interfaces break the momentum conservation of incident particles,
effectively randomizing the momenta of the scattered particles.
Therefore, we approximate the interface coupling as
$ U_{\bm{k},\bm{k}'} \approx U = \textrm{const} $.
Furthermore, we work in the thermodynamic limit where
$ \frac{1}{N_L} \sum_{\bm{k}} = \left( \frac{ a_L }{ 2\pi } \right)^3 \int d^3 k $
and
$ \frac{1}{N_R} \sum_{\bm{k}'} = \left( \frac{ a_R }{ 2\pi } \right)^3 \int d^3 k' $,
and take both leads to be of the same material,
so that we can drop the $L/R$ label.
Then $\tilde{D}^{P/AP}_E(\epsilon) = \tilde{D}^{P/AP}_E(-\epsilon)$,
resulting in $\gamma_{E}^{P/AP} = 0$ [see Eqs.~(\ref{eq:tildeD_P_AP}) and (\ref{eq:gamma_E})];
i.e., there is no additional energy transfer to the lattice.
In keeping with the long-wavelength expansion used in Sec.~\ref{sec:H_X},
we only consider low temperatures $T \ll J S / k_B$.
For yttrium iron garnet \cite{cherepanov1993saga}, this means $T \ll 40\,{\rm K} $.

\begin{figure*}
\includegraphics[width=\textwidth]{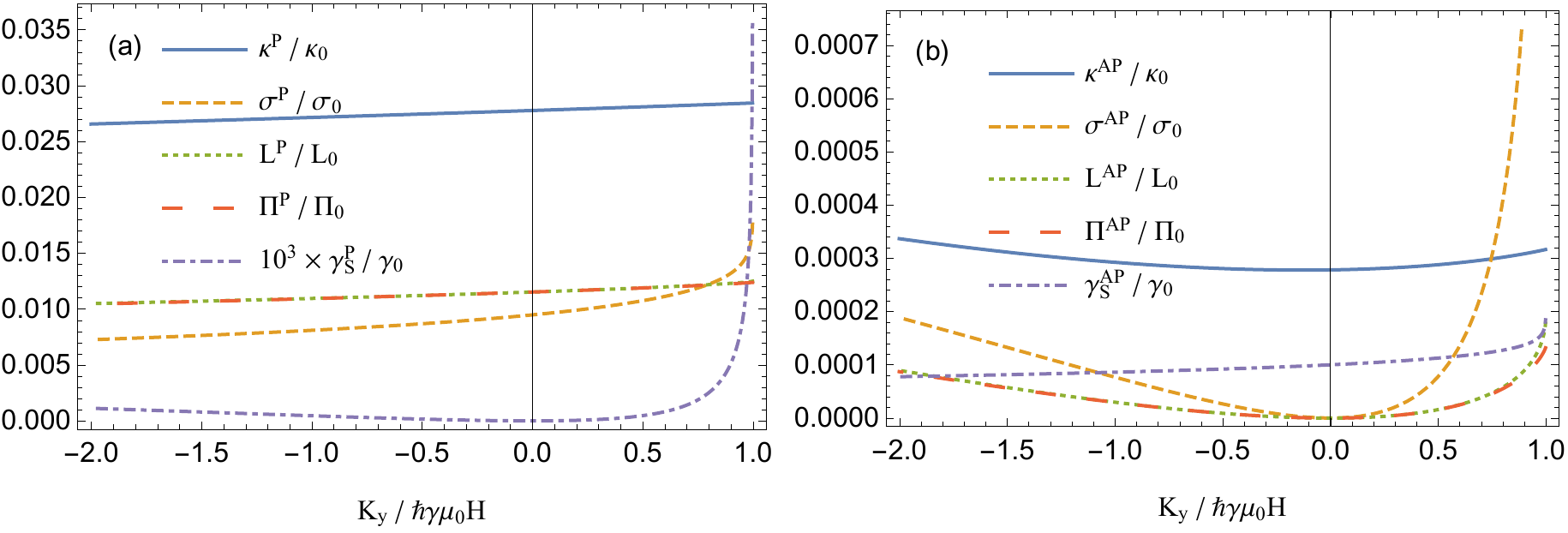}
\caption{ \label{fig:conductance}
Tunneling conductances (\ref{eq:conductances}) in the 
(a) parallel and
(b) antiparallel configurations
as functions of in-plane anisotropy $K_y$,
for $K_x=0$, temperature $k_B T = 10 \times \hbar\gamma \mu_0 H$,
and Gilbert damping parameter $\alpha = 10^{-2}$.
The conductances are rescaled by the dimensionfull prefactors
$ \kappa_0 = U^2 \sqrt{ S k_B^3 T / \hbar^2 J^3 } $,
$ \sigma_0 = \gamma_0 = \hbar \kappa_0 / k_B^2 T $,
$ L_0 = \hbar\kappa_0 / k_b T $, and
$ \Pi_0 = T L_0 / \hbar $.
With this rescaling, 
spin Seebeck and Peltier conductances,  $L/L_0$ and $\Pi/\Pi_0$, respectively, lie almost perfectly on top of each
for most values of anisotropy $K_y$,
reflecting Onsager reciprocity.  
}
\end{figure*}

The tunneling conductances (\ref{eq:conductances}) are displayed in Fig.~\ref{fig:conductance} 
as functions of the in-plane anisotropy, i.e., of the spin-wave ellipticity. 
In the parallel configuration shown in Fig.~\ref{fig:conductance}(a),
all conductances depend only weakly on the magnitude of the anisotropy.
With the exception of the dissipation-assisted spin conductance $\gamma_S^{P}$,
they decrease for hard-axis ($K_y<0$) and increase for easy-plane ($K_y>0$) anisotropy.
This can be attributed to the magnon gap increasing or decreasing, respectively,
which increases or decreases the overall magnon population. 
The strong increase and eventual divergence of the spin conductance for $K_y \to \hbar \gamma \mu_0 H$
signifies the divergence of the Bose distribution for vanishing spin-wave gap,
and is a precursor to the magnetic reorganization transition 
in which the magnetization tilts into the anisotropy plane.
The additional dissipation-assisted spin conductance $\gamma_S^{P}$ mirrors this behavior,	
but also increases for hard-axis anisotropies, in contrast to all other conductances. 
The reason for this is that it is an off-resonance process that is less sensitive to the exact value of the gap 
than the resonant ones, whereas its magnitude is determined by the strength of the anisotropies. 
Because of this, it is also three to four orders of magnitude smaller than the other conductances.
Also, note that the breaking of the Onsager reciprocity  of spin Seebeck and Peltier conductances
by the spin-wave ellipticity and the Gilbert damping is negligible in the parallel configuration. 

In the antiparallel configuration displayed in Fig.~\ref{fig:conductance}(b), on the other hand,
the anisotropy dependence of the conductances is more pronounced.
In agreement with the discussion in Sec.~\ref{sec:currents},
spin and spin Peltier conductances are in this case both zero if there is no anisotropy.
However, chemical-potential driven spin transfer is still possible in this case because
the dissipation-assisted spin conductance $\gamma_S^{AP}$ is finite for $K_y=0$.
Apart from $\gamma_S^{AP}$, which decreases for $K_y<0$, all conductances increase away from $K_y=0$,
although they stay small compared to the conductances in the parallel configuration shown in Fig.~\ref{fig:conductance}.
Note that, as the spin-wave gap closes,
the spin conductance diverges and 
the breaking of Onsager reciprocity becomes visible.

To quantify the effect of Gilbert damping and spin-wave ellipticity on the magnon spin valve,
we introduce 
magnetothermal conductance (MTC), 
magnetospin conductance (MSC),
magneto-Seebeck conductance (MLC), and
magneto-Peltier conductance (MPC) ratios
as follows:
\begin{subequations} \label{eq:ratios}
\begin{align}
{\rm MTC} =& \frac{ \kappa^P - \kappa^{AP} }{ \kappa^P } , \\
{\rm MSC} =& \frac{ \sigma^P + \gamma_S^P - \sigma^{AP} - \gamma_S^{AP} }{ \sigma^P + \gamma_S^P } , \\
{\rm MLC} =& \frac{ L^P - L^{AP} }{ L^P } , \\
{\rm MPC} =& \frac{ \Pi^P + \gamma_E^P - \Pi^{AP} - \gamma_E^{AP} }{ \Pi^P + \gamma_E^P } .
\end{align}
\end{subequations}
In the absence of dissipation and spin-wave ellipticity
there are no currents in the antiparallel configuration,
hence these ratios reduce to $1$.
Their deviation from $1$ thus measures the magnitude 
of dissipation and spin-wave ellipticity effects on the magnon spin valve.
The additional energy and spin currents $\gamma_E^{P/AP}$ and $\gamma_S^{P/AP}$ 
are included in the ratios (\ref{eq:ratios})
because they affect the conductance ratios that will be measured in an experiment,
even though they originate from the lattice and not from the magnons in the other lead.  
\begin{figure*}
\includegraphics[width=\textwidth]{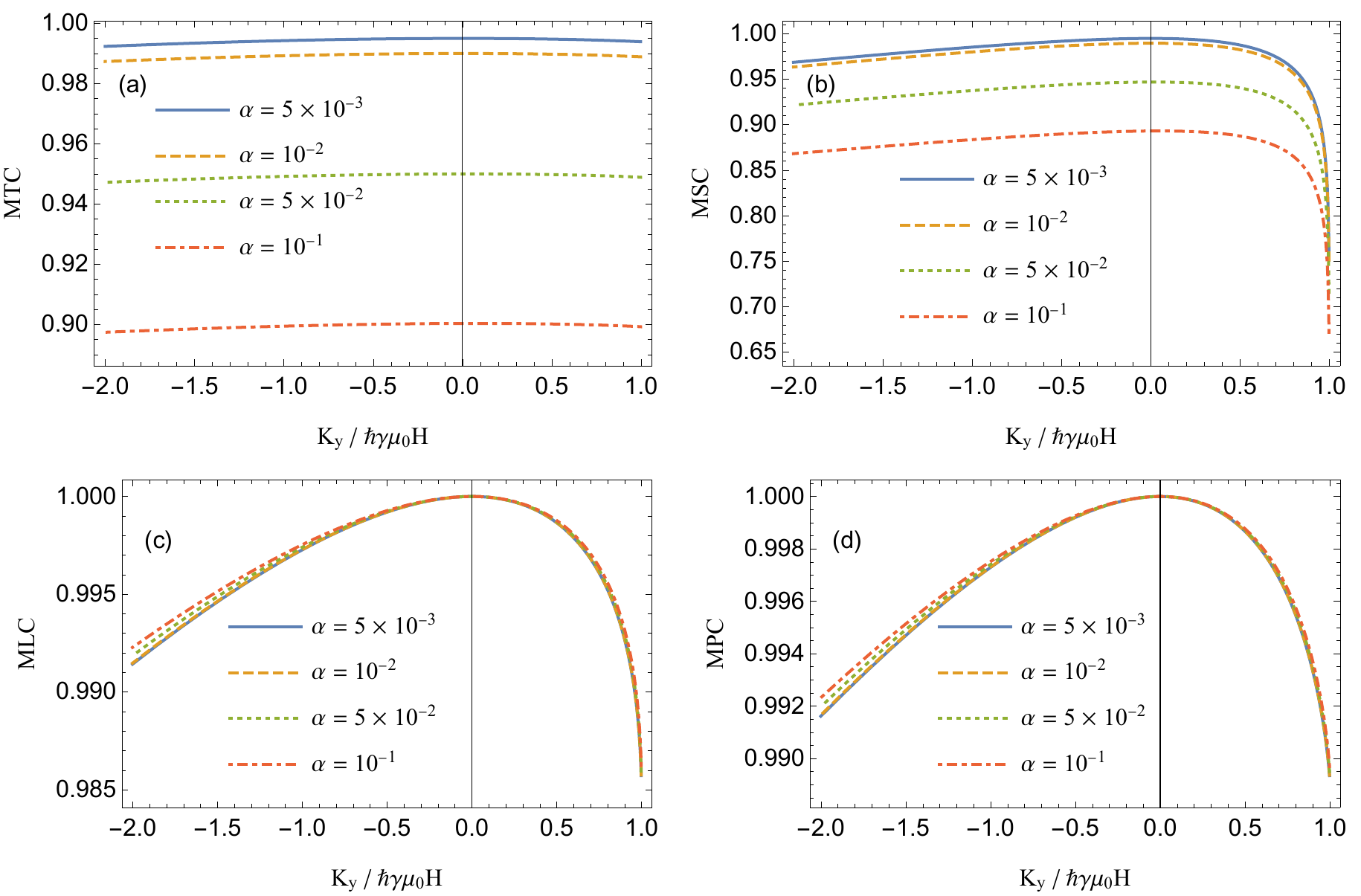}
\caption{ \label{fig:ratios}
(a) Magnetothermal,
(b) magnetospin,
(c) magneto-Seebeck, and
(d) magneto-Peltier conductance ratios as defined in Eqs.~(\ref{eq:ratios}) 
as functions of in-plane anisotropy $K_y$,
for $K_x=0$, temperature $k_B T = 10 \times \hbar\gamma \mu_0 H$,
and various values of the Gilbert damping parameter $\alpha$.
}
\end{figure*}

As shown in Fig.~\ref{fig:ratios}(a),
the in-plane anisotropy affects the MTC ratio only negligibly.
In the presence of large Gilbert damping, on the other hand, the MTC ratio can deviate from $1$ by up to $10\%$.
Responsible for this decrease are the dissipation-assisted pair creation and annihilation processes
that enable energy transfer in the antiparallel configuration even when there is no spin-wave ellipticity. 
The MSC ratio, displayed in Fig.~\ref{fig:ratios}(b), 
behaves similarly for most values of the anisotropy;
when the magnon gap closes for $K_y \to \hbar \gamma \mu_0 H$, however, 
it rapidly decreases because of the divergence of the spin conductance.
 

The magneto-Seebeck and Peltier conductance ratios, 
shown in Figs.~\ref{fig:ratios}(c) and \ref{fig:ratios}(d), respectively,
display an opposite behavior:
they are sensitive to the anisotropy, but only slightly affected by the Gilbert damping.
They show a decrease of the order of $1\%$ when the magnon gap closes for $K_y \to \hbar \gamma \mu_0 H$,
and also decrease, albeit more slowly, for increasing hard axis anisotropy;
this reflects the increasing strength of the spin-conservation breaking in both cases.
Unlike the MTC and MSC ratios, 
the MLC and MPC ratios are actually increased by the Gilbert damping for larger values of the anisotropy.
Note also that the Seebeck and Peltier ratios are both qualitatively and quantitatively almost identical;
the breaking of the Onsager reciprocity is only apparent in the stronger decrease of the MLC ratio for $K_y \to \hbar \gamma \mu_0 H$.

\section{Conclusions}

\label{sec:conclusions}

We have studied the tunneling current in a magnon spin valve device. 
By applying the Holstein-Primakoff transformation to the Heisenberg Hamiltonian, 
we derived the magnon Hamiltonian, 
in which transverse anisotropies introduce ellipticity of the magnons. 
We have also added Gilbert damping 
to the magnon spectral function to study the effects of dissipation on the magnon tunneling.
Both precession ellipticity and Gilbert damping are found to open new,
Onsager-reciprocity breaking channels for heat and spin transport across the junction,
resulting in finite currents even when the magnetizations of both leads are aligned antiparallel.  
We have not only found that dissipation and spin-wave ellipticity decrease
the spin and heat conductance ratios,
but have also revealed a clear difference in the sensitivity of heat and spin currents to these two quantities. 
We hope that our results provide useful guidance for the design and understanding of magnon spin valve devices.

\acknowledgments 

This work is supported by the European Research Council via Consolidator Grant No.~725509 SPINBEYOND. R.D.~is a member of the D-ITP consortium, a program of the Netherlands Organisation for Scientific Research (NWO) that is funded by the Dutch Ministry of Education, Culture and Science (OCW). J.Z.~would like to thank the China Scholarship Council.
This research was supported in part by the National Science Foundation under Grant No.~NSF PHY-1748958.

\appendix

\section*{Appendix: Derivation of the Tunneling Currents}

\label{sec:current_derivation}

In this Appendix, we outline the derivation of the tunneling currents
given in Sec.~\ref{sec:currents}.
The total energy current from the left to the right lead is given by
\begin{align}
I_E^{P/AP} 
&= \partial_t \left\langle {\cal H}_R \right\rangle \\
&= \sum_{\bm{k}'} E_{R,\bm{k}'} \partial_t n_{R,\bm{k}'} ,
\label{eq:I_E_def}
\end{align}
whereas the spin current from the left to right lead is
\begin{align}
I_S^P
&= -\partial_t \sum_j \hbar \left\langle S_{R,j}^z \right\rangle \\
&= \hbar \sum_{\bm{k}'} \left( u_{R,\bm{k}'}^2 + v_{R,\bm{k}'}^2 \right) \partial_t n_{R,\bm{k}'} ,
\label{eq:I_S_P_def}
\end{align}
in the parallel configuration, or
\begin{align}
I_S^{AP}
&= \partial_t \sum_j \hbar \left\langle S_{R,j}^z \right\rangle \\
&= \hbar \sum_{\bm{k}'} \left( u_{R,\bm{k}'}^2 + v_{R,\bm{k}'}^2 \right) \partial_t n_{R,\bm{k}'} .
\label{eq:I_S_AP_def}
\end{align}
in the antiparallel configuration.
Using Fermi's golden rule \cite{bruus2004many}, 
we find the following kinetic equations for the quasiparticle distribution functions
in the parallel configuration:
\begin{widetext}
\begin{subequations} \label{eq:kinetic}
\begin{align}
\partial_t n_{L,\bm{k}} = 
&
\frac{ 2\pi S_L S_R }{ \hbar N_L N_R } \sum_{\bm{k}'} \Bigl[
| U_{\bm{k},-\bm{k}'} |^2 \left( u_{L,\bm{k}} u_{R,\bm{k}'} + v_{L,\bm{k}} v_{R,\bm{k}'} \right)^2
\delta\left( E_{L,\bm{k}} - E_{R,\bm{k}'} \right)  
\left( n_{R,\bm{k}'} - n_{L,\bm{k}} \right)  
\nonumber\\
&
\phantom{ \frac{ 2\pi S_L S_R }{ \hbar N_L N_R } \sum_{\bm{k}'} \Bigl[ }
+
| U_{\bm{k},\bm{k}'} |^2 \left( u_{L,\bm{k}} v_{R,\bm{k}'} + v_{L,\bm{k}} u_{R,\bm{k}'} \right)^2
\delta\left( E_{L,\bm{k}} + E_{R,\bm{k}'} \right)
\left( 1 + n_{L,\bm{k}} + n_{R,\bm{k}'} \right) 
\Bigr] ,
\\
\partial_t n_{R,\bm{k}'} = 
&
\frac{ 2\pi S_L S_R }{ \hbar N_L N_R } \sum_{\bm{k}} \Bigl[
| U_{\bm{k},-\bm{k}'} |^2 \left( u_{L,\bm{k}} u_{R,\bm{k}'} + v_{L,\bm{k}} v_{R,\bm{k}'} \right)^2
\delta\left( E_{L,\bm{k}} - E_{R,\bm{k}'} \right)  
\left( n_{L,\bm{k}} - n_{R,\bm{k}'} \right)  
\nonumber\\
&
\phantom{ \frac{ 2\pi S_L S_R }{ \hbar N_L N_R } \sum_{\bm{k}'} \Bigl[ }
+
| U_{\bm{k},\bm{k}'} |^2 \left( u_{L,\bm{k}} v_{R,\bm{k}'} + v_{L,\bm{k}} u_{R,\bm{k}'} \right)^2
\delta\left( E_{L,\bm{k}} + E_{R,\bm{k}'} \right)
\left( 1 + n_{L,\bm{k}} + n_{R,\bm{k}'} \right) 
\Bigr] .
\end{align}
\end{subequations}
\end{widetext}
The corresponding expressions in the antiparallel configuration can be obtained from Eq.~(\ref{eq:kinetic})
by exchanging the Bogoliubov-coefficient prefactors according to
$ \left( u_{L,\bm{k}} u_{R,\bm{k}'} + v_{L,\bm{k}} v_{R,\bm{k}'} \right)^2
\leftrightarrow
\left( u_{L,\bm{k}} v_{R,\bm{k}'} + v_{L,\bm{k}} u_{R,\bm{k}'} \right)^2 $.
The energy and spin currents (\ref{eq:I_E}) and (\ref{eq:I_S}) are obtained by
inserting the kinetic equations (\ref{eq:kinetic})
into their respective definitions (\ref{eq:I_E_def}) and (\ref{eq:I_S_P_def}) or (\ref{eq:I_S_AP_def}),
assuming a steady state as in Eq.~(\ref{eq:nk_int}),
and broadening the Dirac distributions with finite dissipation according to Eq.~(\ref{eq:dissipation}).

\bibliographystyle{aipnum4-1} 
\bibliography{magnon2018Feb2019}

\end{document}